\documentclass[preprint,12pt]{elsarticle}
\usepackage[T1]{fontenc}
\usepackage{graphicx}
\usepackage{url}
\usepackage{physics}
\usepackage{xcolor}

\begin{document}

\begin{frontmatter}
\title{Coniferest: a {complete} active anomaly detection framework}

\affiliation[sai]{
            organization={Lomonosov Moscow State University, Sternberg Astronomical Institute},
            addressline={Universitetsky pr.~13},
            city={Moscow},
            postcode={119234},
            country={Russia}}

\affiliation[hse]{
            organization={National Research University Higher School of Economics},
            addressline={21/4 Staraya Basmannaya Ulitsa},
            city={Moscow},
            postcode={105066},
            country={Russia}}

\affiliation[ind]{
            organization={Independent researcher},
            }

\affiliation[mcwilliams]{
            organization={McWilliams Center for Cosmology and Astrophysics, Department of Physics, Carnegie Mellon University},
            city={Pittsburgh},
            postcode={PA 15213},
            country={USA}}

\affiliation[urbana]{
            organization={Department of Astronomy, University of Illinois at Urbana-Champaign},
            addressline={1002 West Green Street},
            city={Urbana},
            postcode={IL 61801},
            country={USA}}

\affiliation[clermont]{
            organization={Université Clermont Auvergne, CNRS/IN2P3, LPCA},
            city={Clermont-Ferrand},
            postcode={63000},
            country={France}}

\affiliation[phfmsu]{
            organization={Lomonosov Moscow State University, Faculty of Physics},
            addressline={Leninskie Gory~1-2},
            city={Moscow},
            postcode={119991},
            country={Russia}
}

\affiliation[iki]{
            organization={Space Research Institute of the Russian Academy of Sciences (IKI)},
            addressline={84/32 Profsoyuznaya Street},
            city={Moscow},
            postcode={117997},
            country={Russia}}

\affiliation[surrey]{
            organization={Physics Department, University of Surrey},
            addressline={Stag Hill Campus, GU2 7XH},
            city={Guildford},
            country={UK}}

\author[sai,hse]{M.~V.~Kornilov\corref{cor}}
\cortext[cor]{Corresponding author.}
\ead{kornilov@physics.msu.ru}
\author[ind]{V.~S.~Korolev}
\author[mcwilliams,urbana]{K.~L.~Malanchev}
\author[sai]{A.~D.~Lavrukhina}
\author[clermont]{E.~Russeil}
\author[sai,phfmsu]{T.~A.~Semenikhin}
\author[clermont]{E.~Gangler}
\author[clermont]{E.~E.~O.~Ishida}
\author[clermont]{M.~V.~Pruzhinskaya}
\author[iki,sai]{A.~A.~Volnova}
\author[surrey]{S.~Sreejith}
\author[]{(The SNAD team)}

\begin{abstract}
We present {\tt coniferest}, an open source generic purpose active anomaly detection framework written in Python.  
The package design and implemented algorithms are described. Currently, static outlier detection analysis is supported via the Isolation forest algorithm. Moreover,
Active Anomaly Discovery (AAD) and Pineforest algorithms are available to tackle active anomaly detection problems.
The algorithms and package performance are evaluated on a series of synthetic datasets. We also describe a few  success cases which resulted from applying the package to  real astronomical data in active anomaly detection tasks  within the SNAD project.

\end{abstract}

\end{frontmatter}

\section{Introduction}
\label{intro}

 The ultimate goal of every machine learning (ML) algorithm is to extract information present from large data sets, and in the process, optimise the allocation of human resources devote to it. Ultimately, improving human life. In the context of anomaly detection (AD) strategies this translates into enabling discoveries which would never happen otherwise.  Depending on the considered field, it means that we can discover hidden or upcoming hardware failures, rare software bugs, malicious or fraud activity in software systems, or even new and completely unforeseen astrophysical objects. All these cases have implications either on our understanding of  the Universe as a whole, or on crucial aspects of daily life.


In AD tasks, it is usually assumed that the dataset includes rare objects (or groups of objects) which are very unlikely to exist when working with the hypothesis that all elements in the data set were generate by the same underlying mechanism.
For example, it is often assumed that anomalies are a minority consisting of a few instances, and that they have attribute values that are very different from
those of nominal instances in~\cite{Liu2008}.
In this context, such objects can be caught by one of the many outlier detection algorithms available.
However, in the context of scientific research, both assumptions are not fully correct.
First, being rare is not enough attribute to consider an astrophysical source valuable or scientifically interesting. When outliers are inspected by a human expert, it often happens that most of them are statistically rare but not useful data samples -- since the final expert request is not to find samples in low density parts of the feature space but to solve the real-world problems (e.g. 68 per cent of outliers in \cite{Malanchev2021} were found to be bogus light curves, including issues with survey automatic photometry and external effects such as satelite or plane tracks). Second, sometimes having attribute values slightly above a certain threshold is enough to be considered a scientifically interesting anomaly. For instance, if we imagine a white dwarf with a mass greater than the Chandrasekhar limit, it becomes clear that the mass only needs to be significantly greater within the measurement error, rather than an order of magnitude greater.

However, identifying such cases is far from a trivial task.
Given the volume and complexity of modern data sets,  the expert is usually able to inspect a very small subset of the objects identified as anomalous by traditional algorithms. 
Indeed, inspecting the hardware or checking software behaviour are extremely time consuming tasks, not to mention the specific case of astronomy, where additional measurements may be required in order to make a final decision.  Thus, algorithms able to fully, and optimally,  exploit past  expert decisions when dealing with new and unlabeled data are required.

Active learning (AL) algorithms  constitute a set of learning strategies that employ expert feedback to fine tune the learning model \cite{ALbook}. In the context of AD, one possible approach is to sequentially show to the expert the object with highest anomaly score and using the feedback to update the hyperparameters of the learning model \cite{Das2017}. The ill-posed nature of the AD problem coupled with the AL strategy results in personalized models which are trained to identify, within a large data set, the specific type of anomaly that is interesting to the expert.

While performing research in this field, the SNAD\footnote{\url{https://snad.space}} team noticed still another practical issue preventing the wide spread use of AL strategies in astronomy: the lack of AL algorithms in popular scientific software, like \textsc{scikit-learn}\footnote{\url{https://scikit-learn.org/}}. Recent efforts in this direction are very sparse and do not completely address these issues\footnote{ActiveAD (\url{https://github.com/clarenceluo78/ActiveAD}) hosts many algorithms but documentation is sparse and it is not adapted to astronomical data.}$^{,}$\footnote{Astronomaly~\cite{2021A&C....3600481L}, is posterior to our efforts and follows its own interface design.}. This motivated us to develop our own software framework, as well as completely new algorithms which were specifically designed to the application of active AD strategies in astronomy catalog data. This resulted in a fully automatized environment and optimize algorithms which have fulfilled our expectations throughout the last few years. We believe this tool is now ready to benefit others, so we decided to provide the community with {\tt coniferest}, a ML framework created using our team deep expertise in the field of active anomaly detection. The package accepts as input any rectangular data matrix holding one line per object and one column per feature\footnote{At this moment it does not handle missing data.}. Despite our main interest being on astronomical application, the entire framework is suitable for a large range of scientific applications.

This paper is organized as follows. In Section~\ref{sec:algorithms} we present all currently implemented algorithms employed in the package. In Section~\ref{sec:packge} we highlight package design considerations and overall usage pattern.
In Section~\ref{sec:evaluation} we provide package evaluation and conclusions are given in Section~\ref{sec:conclusion}.

\section{Algorithms}
\label{sec:algorithms}

\subsection{Isolation forest}
\label{sec:isoforest}
Isolation forest (IF) is an outlier detection algorithm described in~\cite{Liu2008} and later in~\cite{Liu2012}.
It is not an active anomaly detection method, however it is the base for other tree-based algorithms implemented within the package.
Each member of the tree ensemble is a particular variant of extremely randomized trees called an iTree~(isolation trees) in the original paper~\cite{Liu2012}.

Considering a binary tree, each node represents a split point and a feature (coordinate index), resulting into the feature space partitioning into non-overlapping multidimensional rectangles.
An arbitrary value can be assigned to every rectangle, for instance a class, for a  classification problem, or a real value, for a regression problem.
Note that it is computationally inexpensive to find the corresponding rectangle for any input point in the feature space. Its  assigned value is then used as the tree prediction for that particular object.
Depending on the problem, the split value and the feature are evaluated for every node in some optimal way to create the optimal feature space partitioning, providing good predictions.

An ensemble of trees --- called as forest --- consists of a number of weak predictors (trees) each solving the same problem, being it a classification, a regression problem, or an outlier isolation problem.
Then, the predictions of every tree are averaged, summed, or taken into account in another manner over the whole forest,  depending on the considered method.
Exactly as Leo Tolstoy writes, \textit{Happy families are all alike; every unhappy family is unhappy in its own way} \cite{tolstoy1899anna}, all trees in the forest provide a slightly wrong prediction in its own way, but using the ensemble averaging allows us to improve prediction quality dramatically.
Usually, every tree needs to be additionally randomized by subsampling the dataset or choosing a subset of considered features,  otherwise all trees would be wrong in a  similar way and the ensemble averaging would not help.
Decision trees for classification via the random forest method make splits in some local-optimal manner, trying to separate samples belonging to different classes as much as possible. In case of AD, isolation trees choose completely random order for both, feature and the split point. At every node a random feature and a random split point, from the domain covered by the data, are chosen.
Empty (or degenerated) splits are prohibited, i.e. every single feature space rectangle encloses at least one sample from the initial dataset.
The single-point rectangle cannot be split further and corresponds to a leaf of the tree.
Then it is fairly obvious that splitting single point into the separate region are more likely when the point is distant from the bulk of the data in the feature space.
Therefore, the regions of feature space populated by outliers appear earlier,  near the root of the tree.
It means that average leaf depths are smaller for the outliers than those for the nominal data.

For every particular point in the feature space it is possible to evaluate the leaf depth in every tree of the forest.
Then, the leaf depths are averaged and used as a measure of abnormality.
Namely, the following anomality score $s(\vb x)$ is used within {\tt coniferest} package,  to be consistent with those used in {\tt scikit-learn}:
\begin{equation}
\label{eq:isolation_forest_score}
s(\vb x) = -2^{-\frac{\bar d(\vb x)}{c(N)}}.
\end{equation}
In the above equation, $c(N)$ is the normalized constant which  depends on the dataset size $N$ following:
\begin{equation}
\label{eq:average_path_length}
c(N) \equiv 2\left(H(N)-1\right) \approx 2 \left(  \gamma - 1 + \ln(N) + \frac{1}{2N} - \frac{1}{12N^2}\right),
\end{equation}
where $H(N)$ is so-called Harmonic number, $\gamma \approx 0.5772$ is Euler's constant and the constant $c(N)$ is essentially an average leaf depth for an isolation tree built for uniformly distributed data. Moreover, $\bar d(\vb x)$ is an ensemble-averaged leaf depth for a given data point,  $\vb x$:
\begin{equation}
\label{eq:mean_path_length}
\bar d(\vb x) \equiv \frac{1}{K} \sum_{i=1}^{K} \left( d\left(l_i(\vb x)\right) + c(N\left(l_i(\vb x)\right)) \right),
\end{equation}
with $K$ representing the number of trees in the forest;
$l_i(\vb x)$ denoting the leaf in $i$-th tree covering point $\vb x$, $d(l)$ being the leaf $l$ depth, and $N(l)$ corresponding to the number of samples from the dataset in leaf $l$.
A popular optimization is employed in our isolation forest implementation: anomaly scores for the nominal data are evaluated only approximately,
since they are not as important as anomaly scores for outliers.
The tree depth is limited to $\log_2(N)$, which allows to speedup the process of building trees.
The term $c(N\left(l(\vb x)\right))$ in Equation \ref{eq:mean_path_length} a  consequence of such optimization.

We notice  that $s(\vb x) \rightarrow -1$ as $\bar d(\vb x) \rightarrow 0$, and $s(\vb x) \rightarrow +1$ as $\bar d(\vb x) \rightarrow +\infty$,
which results in outliers having negative scores.


\subsection{Active Anomaly Discovery}
\label{sec:aad}

Active Anomaly Discovery~(AAD) is a generic active learning technique proposed by~\cite{Das2024,Das2018,Das2017}.
Basically, AAD is not limited by forest algorithms such as Isolation forest.
It is also possible to apply AAD on top of any other outlier detection algorithm.
However, we briefly describe here only the forest flavour of AAD approach, since it is the one currently implemented in the {\tt coniferest} package.

To understand some motivation of this technique, let us note that our final goal is to mimic the decisions provided by a human expert.
If the expert could label the entire dataset, we would have reformulated the problem as a binary classification one.
Then, the Random forest classifier or even Extremely randomized trees could have been applied to solve the problem.
The class (anomaly versus nominal) would have been assigned to every leaf of the classifier.

The above algorithm seems very similar to our previous description of IF, except that a leaf depth, instead of a class, is assigned to every leaf of the forest. Indeed, both IF and Extremely randomized trees, perform a partitioning of the feature space.
The only difference is in the values assigned to the leafs of each tree.
The key idea of AAD is to take IF as
a starting point and then iteratively adapt the leaf values to make the forest similar to a hypothetical Extremely randomized tree classifier, trained to recognise 2 classes (nominal and anomaly), given that the anomaly class is composed by objects that fulfill the expectations of the expert. Considering that IF is a good enough starting point, we can expect that the classifier state is potentially reachable using partially labeled data.

There are two possible sources of partially labeled data.
First, we can optionally provide the algorithm with the labels known in advance, e.g. given well established catalogs or literature reports. Second, the human expert can be asked to input decisions while the active learning loop is running.

For AAD, we have the following equation for the score, which replaces  Equation~(\ref{eq:isolation_forest_score}):
\begin{equation}
\label{eq:aad_forest_score}
s(\vb x) = \sum_{i=1}^{K} w_{i, l_i\left(\vb x\right)} \varphi\left( d\left(l_i(\vb x)\right) + c(N\left(l_i(\vb x)\right)) \right ),
\end{equation}
where $K$ is the number of trees in the forest;
$\varphi(d)$ is some nonlinear transformation function\footnote{Our default choice is $\varphi(d) = -d^{-1}$, however other variants such as $\varphi(d) = -2^{d},$ $\varphi(d) = 1,$ or $\varphi(d) = d$ are also considered~\cite{Das2024,Das2018,Das2017}.} and
$w_{i, l_i\left(\vb x\right)}$ are weights which can be adjusted, allowing  Equation~(\ref{eq:aad_forest_score}) to better predict decisions reported by the human expert.
There are as many different weight parameters as there are leafs in the forest.
For simplicity, we also use notation $w_j$ further each $j$ corresponds to a leaf in the forest.
Additionally, all weights were normalized, so $\lVert \vb w \rVert^2 = 1$.

Considering $\mathcal{A}$ as a known anomaly subset of the data and $\mathcal{N}$ as a known nominal subset, the optimization problem for estimating weights $\vb w$ is defined as:
\begin{multline}
\label{eq:aad_forest_loss}
\vb w = \arg\min_{\vb w} \left(
	\frac{C_a}{\left|\mathcal{A}\right|} \sum_{i \in \mathcal{A}} \mathrm{ReLU}\left(s(\vb x_i | \vb w) - q_{\tau}\right)\right.+\\+
	\left.
	\frac{1}{\left|\mathcal{N}\right|} \sum_{i \in \mathcal{N}} \mathrm{ReLU}\left(q_{\tau} - s(\vb x_i | \vb w)\right) +
	\frac{1}{2} \lVert \vb w - \vb w_0 \rVert^2
	\right).
\end{multline}
The first term in Equation (\ref{eq:aad_forest_loss}) is a loss due to known anomalies predicted as nominals by the model.
The second term is a loss due to known nominals wrongly predicted as anomalies and
the third term accounts for regularization.
The constant $C_a$ allows us to assign a relative importance of to both kind of errors, our current default choice being  $C_a=1$.
The threshold score $q_{\tau}$ is the limiting score necessary to identify the $1-\tau$ percentile of the dataset. Our current default choice is $\tau = 0.97$.
This means that all known anomalies should occupy the top 3\% anomaly scores, which is achieved by adjusting the weights $\vb w$.
The $\vb w_0$ in regularization term is either an uniform vector or the weights from the previous iteration, while
$\mathrm{ReLU}$ denotes the rectified linear function:
\begin{equation}
\mathrm{ReLU}(x) \equiv \begin{cases}
x\quad x \ge 0,\\
0\quad\mathrm{otherwise}.
\end{cases}
\end{equation}
Solving the optimization problem described by Equation (\ref{eq:aad_forest_loss}) is challenging due to high number of unknown parameters to be determined.
Currently, we use {\tt trust-krylov} method from {\tt scipy} optimization framework, which uses the Newton GLTR trust-region algorithm~\cite{Gould1999}.

At each iteration, Equation ~(\ref{eq:aad_forest_loss}) is solved and the  scores for the entire dataset are reevaluated using Equation ~(\ref{eq:aad_forest_score}) and the new weights $\vb w$. Thus, resulting in a new unknown object associated to the highest anomaly score and a new $\tau$ score quantile.
The most anomalous unknown object is shown to the human expert.
Depending on the expert decision, this object is added to either $\mathcal{A}$ or $\mathcal{N}$, and the next iteration occurs.
We emphasize that it is crucial to perform the model inference as fast as possible,  since estimations for the entire dataset are reevaluated at every iteration.

The total number of iterations, called budget, is an input chosen by the expert, and should be determined taking into account the available time, as well as other resources, necessary to properly judge the candidates presented by the algorithm.
Once the budget is exhausted, we can calculate one of the most important metrics used to evaluate method effectiveness: tje number of expert-approved anomalies within the budget.
In case of simulated data, 
we can also use convenient metrics, e.g.   confusion matrices. However, this is not possible in a real data scenario since the user would not know how many interesting anomalies are included in the entire data set. 

\subsection{Pineforest}
\label{sec:pineforest}

Pineforest\footnote{Korolev et al., in prep.} presents another approach of  refining IF by incorporating feedback from experts. Unlike the previous method, Pineforest does not directly modify existing trees. Instead, it selectively discards trees that inadequately represent the underlying data distribution.

In order to determine the suitability of each tree, a scoring mechanism is devised based on the labeled data. We assign scores to the data points as follows:

\[
y_i = \begin{cases}
-1 & \text{if }\vb x_i\text{ is labeled as an anomaly,} \\
0 & \text{if }\vb x_i \text{ is unlabeled,} \\
1 & \text{if }\vb x_i\text{ is labeled as a regular point.}
\end{cases}
\]
For any given tree \(\vb t_i\), we calculate its score as:

\[
s(\mathbf{t}_i) = \sum\limits_{j=1}^{N} y_j \cdot d(l_i(\vb x_j)),
\]
where \(d(l_i(\ldots))\) denotes the depth of the sample \(\vb x_j\) in tree \(\vb t_i\).

Using these scores, we adapt the forest to the labeled data by discarding trees with lower scores and retaining those with higher ones. This iterative process involves building an initial set of trees, filtering out a certain percentage of them based on their scores (e.g., discarding 90\%), and subsequently rebuilding the forest with the remaining trees. This procedure can be repeated multiple times for further refinement.

The scoring mechanism is designed to prioritize trees that accurately capture regular points deep within the forest, while potentially isolating anomalies at shallower depths. Following the acquisition of new data, the learning process can be reiterated to incorporate the updated information, thus continually adapting to evolving data distributions.

\section{{\tt coniferest} package}
\label{sec:packge}

The {\tt coniferest} package is an open source software, and we follow all best practices currently adopted for developing such tools.
The source codes are available at GitHub: \url{https://github.com/snad-space/coniferest} according the terms of MIT license.

The package is mainly written in Python, since this language is \textit{de-facto}  standard within contemporary academic machine learning communities.
However,  the performance-critical parts, such as tree search algorithms,   required the use of native binary code.
We choose to use Cython to write the performance-critical code, which is latter compiled to native binary code.
Cython is a famous solution for such cases, for instance {\tt scikit-learn} is heavily related on Cython.
It is also popular in the machine learning community, which simplifies possible contributions from the community.
Using Cython allows us to support parallel and fast IF inference, which is required by  the AAD algorithm.
In alternative to Cython, performance-critical code could be written in C++ or Rust.

Using native compiled binary code produced by Cython, C++, or Rust complicates the package installation, particularly for non-experienced users.
In order to circumvent this issue, we provide pre-built python packages in the Wheel format, distributed via the PYthon Package index, for all major platforms, including Windows (both 32-bit and 64-bit architectures), MacOS (both Intel and ARM processors), and Linux (x86\_64 and arm64 are supported).
It means that {\tt pip install coniferest} easily does the job for most users.
The pre-built packages are compiled in the GitHub Actions environment automatically, which simplifies package preparation and makes it more reproducible, reliable and secure for the end user. GitHub Actions are also used to build the code and run unit tests on different Python versions and operation systems.
Employing unit testing helps finding  regressions and dramatically improves overall code quality.

There are only a few dependencies for {\tt coniferest}.
Obviously, {\tt numpy} is on the list, and we also currently re-use {\tt scikit-learn} for building trees.
The latter is a temporarily solution, because currently we have to use internal {\tt scikit-learn} interface which is subject to change without notice.
Using internal interfaces of other libraries dramatically complicates development and distribution of the package, since it requires carefully handling dependencies  versions.
Additionally, we would like to support parallel (yet reproducible) tree building in future versions of our package, which may come in hand for Pineforest algorithm.

We emphasize that, while IF algorithm is implemented within {\tt scikit-learn} package, we had to re-implement it to enhance its performance.
While training is typically the most time-consuming aspect of most machine learning algorithms, for IF and active learning strategies based upon it, scoring is the more demanding process.
This is because scoring, which is essential for anomaly detection, requires the use of the entire dataset, whereas training often only uses a small subset of the original data.

In order to provide useful documentation for the package, we use Python doc-string, to describe functions and classes, and Markdown,  for long read tutorials.
Both are rendered automatically by ReadTheDocs service and available online at \url{https://coniferest.snad.space}.

Even though active anomaly detection usually assumes some active session and the final product are filtered subsets of the input data,
we also provide support for serializing the trained models to portable ONNX format.
This is useful for re-using the trained models for automatic anomaly detection pipelines.
Under such circumstances, the model cannot be update further, but it can be already satisfactory pre-trained by the expert to produce a reasonable stream of data.

\subsection{Models and the {\tt Session}}

There are three basic kinds of entities in {\tt coniferest} packages: datasets, models, and sessions.
The datasets are used solely for testing and demonstrating purposes.
They are not required in production, since each user can analyze their own data.

The models are {\tt scikit-learn}-style classes, implementing every supported algorithm: Isolation forest (IF), Active Anomaly Discovery (AAD) and Pineforest.
These classes mimic the well-known {\tt scikit-learn} interface with a few exceptions, because {\tt scikit-learn} has poor support for partially-labeled data, and consequently there was no good interface to follow.
However, for the {\tt scikit-learn} user it will not be difficult to get acquainted with the {\tt coniferest} interface.

The sessions are objects to support iterative training of the models, as it is done in active anomaly detection. The idea behind the session is to create glue layer between the human expert and the retraining of the model.
Unlike {\tt scikit-learn} model class, the {\tt Session} class constructor takes data and metadata at the object construction time.
The data are ordinary two-dimensional array of the features.
The metadata are one-dimensional array of auxiliary data, only  required to help the  human expert in distinguishing the objects inside main dataset.
In the simplest case the sample metadata are their row index, however, a more description choice is also possible which would help identifying each sample.

The session constructor also requires the existing model object to be created, which  can either be pre-trained or not.
Currently, both AAD and Pineforest models are supported.

The end user can modify the session behaviour by providing callbacks to the object constructor.
The callbacks are used to input the expert decision, to visualize the learning process, to snapshot current model and to save results, among other tasks.
For instance, in our workflow, we supply the expert with a SNAD viewer~\cite{Malanchev2023} URL, pointing to  a portal containing extensive information about the object under consideration. The code then expects a decision (anomaly or not anomaly) to be input interactively.

An alternative to callbacks could have been to use an abstract class and inheritance to tune desired session behaviour.
However, we found that it could be difficult for a normal data scientist to grasp the object-oriented concept. So, we decided to use the functional approach.

\subsection{Supplied datasets}

In order to make the package self-consistent some toy datasets are also supplied within the code.
It is a modern good custom in the machine learning community to provide the dataset within the code package in order to allow the user to start exploring the framework immediately. Currently, we provide the following datasets.

In documentation and tutorial one can find references to an astronomical dataset {\tt ztf\_m31}.
It is a sky field around the M31 galaxy from Zwicky Transient Facility\footnote{\url{https://www.ztf.caltech.edu/}} (ZTF) \cite{2019PASP..131a8002B} survey with features extracted as described in ~\cite{Malanchev2021}.
For those who don't like astronomy, we adopted datasets collected by~\cite{Pang2019} for their work in neural network anomaly detection ({\tt dev\_net\_dataset}).
The following datasets are available:
{\tt donors}~(school projects excitement), {\tt census}~(high-income person), {\tt fraud}~(credit card transactions), {\tt celeba}~(annotated celebrity images), {\tt backdoor}~(backdoor attack data), {\tt campaign}~(phone calls marketing), {\tt thyroid}~(disease detection data set).
For further detailed description, please see~\cite{Pang2019}.

There is also a debugging dataset called {\tt non\_anomalous\_outliers}, this dataset is generated by request and consists of nominal data and a required fraction of outliers.

\section{Evaluation}
\label{sec:evaluation}

\subsection{Isolation forest prediction}

\begin{figure}
\centering
\includegraphics[width=8cm]{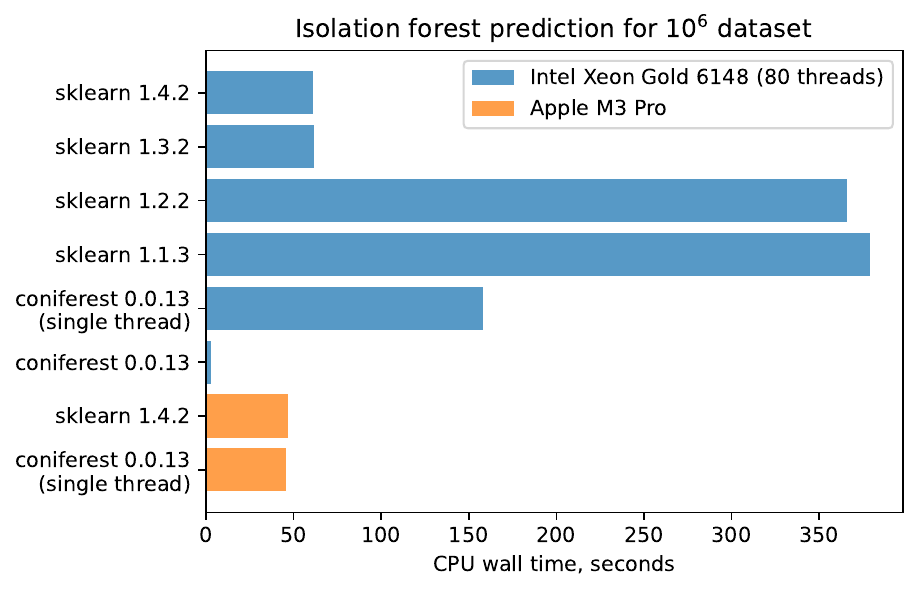}
\caption{
\label{fig:isoforest}
Performance comparison between {\tt scikit-learn} (listed as "sklearn") and {\tt coniferest}.
A dataset containing $\approx 10^6$ samples,  each having $2$ features is considered.
The dataset anomaly score evaluation has been measured for different versions of {\tt scikit-learn}.
Additionally, {\tt coniferest} in single-thread mode is considered.
Note, that {\tt scikit-learn} is always single-threaded.
}
\end{figure}

As we noted before, it is important to be able to quickly evaluate the model for AAD  and Pineforest.
Unfortunately, at the time of the creation of {\tt coniferest}, IF prediction within {\tt scikit-learn} was single-threaded and not quite optimized, even in single-threaded mode.
While {\tt scikit-learn} performance has been sufficiently improved later, it is still impossible to make multi-threading predictions.
There is two ways to make multi-threading prediction for IF: process each tree in a separate thread or process each data sample in a separate thread.
We use the latter approach.
Performance comparison is shown in  Figure ~\ref{fig:isoforest}.

\subsection{Success cases}

In~\cite{Volnova2023}, {\tt coniferest} package has been employed within ZWAD pipeline~\cite{Malanchev2021a} to perform anomaly detection for ZTF DR17 dataset.
The human expert was happy to find anything astrophysically exciting.
For instance, a unknown binary microlensing event AT~2021uey has been found, an optical counterpart for radio source NVSS
J080730+755017, and lots of variable stars and fast transients, such as red dwarf flares.

The same ZWAD pipeline has been used in~\cite{Pruzhinskaya2023}.
Instead of general anomaly detection problem, the problem was stated as semi-supervised classification one.
This allowed us to discover 104 supernovae (including 57 previously unreported) in ZTF DR3 dataset.
The human expert was serching only  for supernovae,  and answered to the active anomaly algorithm accordingly.
About 2000 candidates were inspected, and labeled, leading to the discovery of 100 supernovae.
Similarly, red dwarf flares were of interest in~\cite{2024MNRAS.tmp.1989V}.
ZTF DR8 dataset was used to find red dwarf flare events and about 130 flares were found within 1200 candidates.

Note that instead of inspecting millions of light curves in ZTF dataset only tiny subsets were scrutinized while looking for supernovae and red dwarf flares.
From astrophysical intuition we understand that it is unlikely to one would find 100 supernovae within random 2000 samples subset of ZTF dataset,
otherwise it would mean that the entire Galaxy is consisted of supernovae which is obviously not true.

\section{Conclusion}
\label{sec:conclusion}

We introduced the {\tt coniferest} package, a multi-algorithm and open-source software designed for anomaly detection, developed by the SNAD team. The software, written in Python, supports a range of algorithms, including Isolation Forest (IF) for static anomaly detection,  Active Anomaly Discovery (AAD) and Pineforest for active anomaly detection. The framework’s integration of Cython for performance-critical tasks ensures efficient parallel processing, enhancing the speed and scalability of anomaly detection tasks. Coniferest’s design aligns with modern machine learning practices, offering a user-friendly interface similar to {\tt scikit-learn}, and supports the serialization of trained models in ONNX format for seamless integration into automated pipelines.

The successful application of {\tt coniferest} algorithms for the Zwicky Transient Facility data highlights its effectiveness in identifying astrophysically significant events such as binary microlensing, optical counterparts for radio sources, rare variable stars, and supernovae. This framework not only simplifies the implementation of anomaly detection algorithms, but also bridges the gap between data scientists and domain experts, fostering more effective and efficient identification of anomalies in large datasets. As we continue to develop and refine {\tt coniferest}, we anticipate its broader adoption across different fields, facilitating more accurate and timely detection of anomalies in diverse datasets, which will be particularly beneficial for the V. Rubin Observatory Legacy Survey of Space and Time\footnote{\url{https://www.lsst.org/}}.

\section*{Acknowledgements}
M.~Kornilov, A.~Lavrukhina, A.~Volnova and T.~Semenikhin acknowledges support from a Russian Science Foundation grant 24-22-00233, \url{https://rscf.ru/en/project/24-22-00233/}. Support was provided by Schmidt Sciences, LLC. for K.~Malanchev.

\section*{Disclosure of Interests}
The authors declare no conflicts of interest.

\bibliographystyle{elsarticle-num}
\bibliography{refs}
\end{document}